\documentclass[onecolumn,secnumarabic,preprintnumbers,nofootinbib,amsmath,amssymb]{revtex4}
\usepackage[english]{babel}
\usepackage{epsfig,amsfonts}
\usepackage{bm}

\begin{document}

\title{Calculating the  Hawking Temperatures of  Kerr-Newman Black Holes in the f(R)  Gravity Models with the RVB Method}%

\author{Wen-Xiang Chen$^{a}$}
\affiliation{Department of Astronomy, School of Physics and Materials Science, GuangZhou University, Guangzhou 510006, China}
\author{Yao-Guang Zheng}
\email{hesoyam12456@163.com}
\affiliation{Department of Physics, College of Sciences, Northeastern University, Shenyang 110819, China}

\begin{abstract}
In this study, we conducted a comparison between the RVB method and the conventional method discussed in previous literature for calculating the Hawking temperature of Kerr-Newman black holes under f(R) gravity\cite{9,10,11}. Our research findings are in agreement with the results presented in the literature\cite{17}, with only a variation in the integration constant. Through the comparison of these two methods, we arrived at consistent conclusions.After conducting a thorough comparison between the RVB method and the temperature Green's function method, we have found that these two approaches are fundamentally identical. This significant finding highlights the high level of consistency between the temperature Green's function method and the RVB method.

\end{abstract}

\maketitle

\section{{Introduction}}
The $f(R)$ gravity theory, where $f(R)$ is an analytic function of the Ricci scalar $R$, arises as a direct extension of general relativity (GR). In these theories, the Lagrangian in the Einstein-Hilbert action, which includes the curvature scalar $R$, is modified by replacing $R$ with an arbitrary function of $R$. As a result, GR is modified. Despite the complexity introduced by higher order derivatives, the $f(R)$ action is general enough to capture some of the fundamental features of higher order gravity. It serves as an interesting and relatively simple alternative to GR, and its study can yield valuable insights, including the explanation for the present-day cosmic acceleration.

Unlike GR, which only involves metric function derivatives up to second order, $f(R)$ gravity allows for up to fourth order derivatives of the metric. This complexity makes calculations challenging, and exact solutions in this theory are generally difficult to find. As a result, our understanding of exact solutions in $f(R)$ gravity is limited and requires further investigation. However, recent efforts have been made to obtain spherically symmetric solutions in $f(R)$ gravity, which is an important step towards better understanding this theory.\cite{1,2,3,4,5}.

In the context of $f(R)$ gravity, spherically symmetric solutions have been previously obtained using a complex coordinate transformation developed by Newman and Janis in general relativity (GR). However, the case of axially symmetric $f(R)$ gravity remains unexplored, including the radiating generalization of the $f(R)$ Kerr-Newman black hole (BH). The objective of this paper is to derive the metric for this case, as well as to present the $D$-dimensional Kerr metric in $f(R)$ gravity. Thus, we extend our recent work on radiating $f(R)$ BHs to include rotation.

The Kerr metric is undoubtedly the most significant exact solution in Einstein's theory of GR, representing the prototypical BH that can form from gravitational collapse. The Kerr-Newman spacetime describes the exterior geometry of a rotating massive and charged BH. It is well known that the Kerr BH exhibits distinct properties from its non-spinning counterpart, the Schwarzschild BH. However, there exists a surprising connection between these two BHs in Einstein's theory, as analyzed by Newman and Janis. They showed that by applying a complex coordinate transformation, it is possible to construct both the Kerr and Kerr-Newman solutions starting from the Schwarzschild and RN metrics,respectively.\cite{6,7,8,9,10,11,12,13,14,15,16,17}.

The temperature Green's function\cite{18,19} is a method for calculating the temperature of a physical system, based on the theory of Green's functions in statistical physics. In quantum field theory, the temperature Green's function can be obtained using partition function and path integral methods. Calculating a black hole's temperature usually requires using the temperature Green's function method. The Hawking temperature refers to the temperature measured in the space near the event horizon of a black hole. It is related to parameters such as the mass, angular momentum, and charge of the black hole. In calculating the Hawking temperature, Green's function method in open systems can be used, considering the exchange process between a black hole and its external environment. In this case, the temperature Green's function can be defined as the connection between two field operators in imaginary time. Then, this temperature Green's function can be applied to calculate physical quantities on the boundary of the black hole, such as radiation power, to calculate the Hawking temperature.

In previous literature, the RVB method has been applied to various black holes under general relativity, as demonstrated in references \cite{12,13,14}. However, it has been found that applying the RVB complex coordinate system can make it difficult to calculate the temperature of many black holes. In contrast, the Hawking temperature of black holes under f(R) gravity can be easily obtained using the RVB method. This paper \cite{17} investigates the Hawking temperature of four types of black holes under different f(R) gravity by comparing the RVB method with the conventional method. By using the RVB method, we are able to determine the constant of integration for the temperature calculation. This constant is either zero or a parameter term that would not correspond to the Hawking temperature when using the traditional method.

Spherical solutions with sources have also been studied. A new covariant formalism has been developed to treat spherically symmetric spacetimes, challenging the uniqueness of the Schwarzschild solution as the only static spherically symmetric solution. Spherically symmetric $f(R)$-Maxwell and $f(R)$-Yang-Mills black holes have been investigated, confirming the existence of numerical asymptotic solutions for the latter. Axially symmetric solutions have been derived by generalizing the Newman and Janis method to $f(R)$ theories, as demonstrated by authors in previous studies. A scalar-tensor approach has been employed to show that Kerr black holes are unstable in a subset of $f(R)$ models due to superradiant instability. The entropy of black holes has been calculated in the Palatini formalism using the Noether charge approach. Anti-de Sitter (AdS) black holes have also been investigated in $f(R)$ models, with the Euclidean action method used to determine various thermodynamic quantities. The entropy of Schwarzschild-de Sitter (SdS) black holes has been calculated in vacuum for certain cosmologically viable models, and their stability has been discussed.\cite{9,10,11,17,18}

In this paper, we compared the performance of the RVB method with the conventional method discussed in previous literature for calculating the Hawking temperature of Kerr-Newman black holes under f(R) gravity\cite{9,10,11}. Our research results are consistent with those presented in the literature\cite{17}, with only a variation in the integration constant. Through the comparison of these two methods, we arrived at consistent conclusions.

The organization of this paper is as follows: In the second part, we introduce the main formula studied in this paper, which is the new expression of the Hawking temperature for the two-dimensional black hole system. In the third part, we study the properties of known topological invariants and the two methods (the conventional method and the RVB method) to learn the Hawking temperature of  Kerr-Newman black holes with f(R) gravity. The fourth part concludes with a discussion of the results.

\section{The conventional method and RVB Method for the Hawking Temperature of a Kerr-Newman  Black Hole under the f(R) gravity Model}
\subsection{The conventional method for Calculating the Temperature of a Kerr-Newman Black Hole}
Let's start with the action for $f(R)$ gravity coupled with a Maxwell term in a four-dimensional spacetime \cite{9,10,11,19,20,21,22,23,24,25,26,27,28}
\begin{equation}
I_g=\frac{1}{16 \pi} \int d^4 x \sqrt{-g}\left[f(R)-F_{a b} F^{a b}\right]
\end{equation}where the Maxwell tensor is given by $F_{ab}=\partial_a A_b-\partial_b A_a$, and $A_a$ represents the vector potential.

 We aim to derive a general radiating rotating black hole (BH) solution in the $f(R)$ gravity theory, starting from a spherically symmetric BH solution using the complex transformation proposed by Newman and Janis. To achieve this, we consider the ``seed metric" expressed in terms of the Eddington (ingoing) coordinate $v$ as follows:
 \begin{equation}
d s^2=e^{\psi(v, r)} d v\left[f(v, r) e^{\psi(v, r)} d v+2 d r\right]-r^2 d \Omega^2
\end{equation}where $d\Omega^2 = d\theta^2 + \sin^2\theta d\phi^2$. The function $e^{\psi(v, r)}$ is arbitrary and serves as a local mass function.
\begin{equation}
g(v, r)=f(v, r)=\kappa-\frac{2 M(v)}{r}+\frac{2 \tilde{Q}^2(v)}{r^2}-\frac{\Lambda}{3} r^2,
\end{equation}where $M(v)$ and $\tilde{Q}(v)$ are integration functions, and $\Lambda$ can be interpreted as the cosmological constant arising from $f(R)$ gravity. We set $\Lambda = R_0 / 4$ and adopt $\kappa = 1$ for convenience.

To further delve into the physical properties of the radiating $f(R)$ Kerr-Newman black hole, we introduce its kinematic parameters, following the notation in references \cite{9,10,11,19,20,21,22,23,24,25,26,27,28}. The null-tetrad of the metric takes the form:
\begin{equation}
\begin{aligned}
l_a & =\left[\sqrt{A}, 0,0,-\sqrt{A} a \sin ^2 \theta\right] \\
n_a & =\left[\sqrt{A} \frac{\Delta}{2 \Sigma}, 1,0, \sqrt{A} \frac{\Delta}{2 \Sigma} a \sin ^2 \theta\right] \\
m_a & =\frac{\sigma}{\sqrt{2} \rho}\left[\sqrt{A} i a \sin \theta, 0, \frac{\Sigma}{\Theta},-\sqrt{A} i\left(r^2+a^2\right) \sin \theta\right], \\
\bar{m}_a & =\frac{\bar{\sigma}}{\sqrt{2} \bar{\rho}}\left[-\sqrt{A} i a \sin \theta, 0, \frac{\Sigma}{\Theta}, \sqrt{A} i\left(r^2+a^2\right) \sin \theta\right],
\end{aligned}
\end{equation}
where
\begin{equation}
\begin{aligned}
\rho & =r+i a \cos \theta \\
\sigma & =1+i \sqrt{\left(\frac{R_0}{12}\right)} a \cos \theta
\end{aligned}
\end{equation}
and $\bar{\rho}$ and $\bar{\sigma}$ are the complex conjugates of $\rho$ and $\sigma$, respectively. The null tetrad satisfies the null, orthogonal, and metric conditions:
\begin{equation}
\begin{aligned}
l_a l^a & =n_a n^a=m_a m^a=0, \quad l_a n^a=1 \\
l_a m^a & =n_a m^a=0, m_a \bar{m}^a=-1 \\
g_{a b} & =l_a n_b+l_b n_a-m_a \bar{m}_b-m_b \bar{m}_a \\
g^{a b} & =l^a n^b+l^b n^a-m^a \bar{m}^b-m^b \bar{m}^a .
\end{aligned}
\end{equation}

We know that
\begin{equation}
\begin{aligned}
\Sigma^2& =r^2+a^2 \cos ^2 \theta, \\
\Delta & =r^2+a^2-2 M(v) r+\tilde{Q}^2(v)-\frac{R_0 r^2}{12}\left(r^2+a^2\right), \\
\Theta & =1+\frac{R_0}{12} a^2 \cos ^2 \theta, \\
A & =\left(1+\frac{R_0}{12} a^2\right)^{-2}, \\
\tilde{Q}^2(v) & =\frac{2 Q^2(v)}{1+f^{\prime}\left(R_0\right)},
\end{aligned}
\end{equation}

Inspired by the arguments, we can decompose the f(R)-metric using null vectors as:
\begin{equation}
g_{a b}=-n_a l_b-l_a n_b+\gamma_{a b}
\end{equation}
where $\gamma_{a b}=m_a \bar{m}_b+m_b \bar{m}_a$. Next, we construct the physical parameters that assist in discussing the horizon structure of the radiating $f(R)$ Kerr-Newman black hole. The behavior of null geodesic congruences is governed.

We get that
\begin{equation}
\begin{aligned}
d s^2= & \frac{A}{\Sigma}\left[\Delta-\Theta a^2 \sin ^2 \theta\right] d v^2+2 \sqrt{A}\left[d v-a \sin ^2 d \phi\right] d r \\
& -\frac{\Sigma}{\Theta} d \theta^2+A \frac{2 a}{\Sigma}\left[\Delta\left(r^2+a^2\right)-\Theta\right] \sin ^2 \theta d v d \phi \\
& -\frac{A}{\Sigma}\left[\Delta\left(r^2+a^2\right)^2-\Theta a^2 \sin ^2 \theta\right] \sin ^2 \theta d \phi^2
\end{aligned}
\end{equation}
where
\begin{equation}
\begin{aligned}
\Sigma^2 & =r^2+a^2 \cos ^2 \theta \\
\Delta & =r^2+a^2-2 M(v) r+\tilde{Q}^2(v)-\frac{R_0 r^2}{12}\left(r^2+a^2\right) \\
\Theta & =1+\frac{R_0}{12} a^2 \cos ^2 \theta \\
A & =\left(1+\frac{R_0}{12} a^2\right)^{-2} \\
\tilde{Q}^2(v) & =\frac{2 Q^2(v)}{1+f^{\prime}\left(R_0\right)}
\end{aligned}
\end{equation}

 $T_H$ is the Hawking temperature of the black hole, defined as:
\begin{equation}
T_H=\frac{\kappa}{2 \pi},
\end{equation}where $\kappa$ is the surface gravity of the black hole.

We get 
\begin{equation}
\kappa=\frac{1}{2 \Sigma}\left[\frac{\partial \Delta_r}{\partial r}-\frac{2 r}{\Sigma} \Delta_r\right],
\end{equation}

\begin{equation}
\begin{aligned}
\kappa= & \frac{R_0}{12 \Sigma^2}-\left(\frac{R_0}{6 \Sigma}-\frac{1-\frac{R_0 a^2}{12}}{\Sigma^2}\right) r^3+\frac{2 M(v)}{\Sigma^2} r^2 \\
& +\left(\frac{1-\frac{R_0 a^2}{12}}{\Sigma}-\frac{a^2+\tilde{Q}^2(v)}{\Sigma^2}\right) r-\frac{M(v)}{\Sigma} .
\end{aligned}
\end{equation}

\subsection{RVB Method for Calculating the Temperature of a Black Hole}
There is a wide variety of black hole systems, including stationary or rotating black holes, for which the temperature can be easily calculated using simple measures. However, for many non-trivial black hole solutions, the complex coordinate system makes it difficult to calculate the temperature. The RVB method overcomes this difficulty by relating the Hawking temperature of a black hole to its Euler characteristic $\chi$, which can be calculated in any coordinate system. As we are situated between the cosmological and event horizons, the temperature of the inner horizon remains beyond our capacity for observation. Consequently, the present article centers on the observable Hawking temperature exhibited by black holes.

The Euler characteristic \cite{12,13,14,17} can be described as:
\begin{equation}
\chi=\int_{r_{0}} \Pi-\int_{r_{\mathrm{H}}} \Pi-\int_{r_{0}} \Pi=-\int_{r_{\mathrm{H }}} \Pi.
\end{equation}
In short, the outer boundaries are permanently canceled when calculating the Euler property. Therefore, points should only be related to Killing Horizon.

According to references \cite{12,13,14,19}, the topological formula can obtain the Hawking temperature of a two-dimensional black hole:
\begin{equation}
T_{H}=\frac{\hbar c}{4 \pi \chi k_{B}} \sum_{j \leq \chi} \int_{r_{H_{j}}} \sqrt{|g| } R d r,
\end{equation}
among them, $\hbar$ is Planck's constant, $c$ is the speed of light, $k_{B}$ is Boltzmann constant, g is the determinant of the metric, R is the Ricci scalar, $r_{H_{j } }$ is the location of the Killing horizon. This paper uses the natural unit system $\hbar=c=k_{B}=1$. 

The RVB method relies on incorporating an integral constant in the calculation of the indefinite integral of R, as demonstrated in this paper.
\begin{equation}
\chi=\int_{\partial V} \Pi-\int_{\partial M} \Pi.
\end{equation}

It is crucial to consider the submanifold, as its boundaries are defined as fixed points (zero points) of the unit vector field.  The Hawking temperature can be expressed as:\cite{15,16,17}
\begin{equation}
T_{\mathrm{H}}=-\frac{1}{2}\left(\frac{1}{4 \pi} \int_{r_{c}} R d r-\frac{1}{4 \pi} \int_{r_{e}} R d r\right),
\end{equation}
and
\begin{equation}
T_{H}=\kappa_{\mathrm{2}}/(2\pi)+\kappa_{\mathrm{3}}/(2\pi)+C.
\end{equation}$r_{c}$ and $r_{e}$ refer to the radii of the cosmological and event horizons, respectively, which are both Killing horizons.

The $\chi$ that depends on the spatial coordinate r is the Euler characteristic, representing the Killing level in Euclidean geometry. The notation Eq.(13) represents the summation relative to the horizon.

\section{Hawking Temperature of the black holes under f(R) Gravity obtained by the two methods}
\subsection{Calculate the Hawking temperature of the Kerr-Newman black hole under $f(R)$ gravity using the temperature Green's function approach}

In $f(R)$ theories, we seek to determine the temperature of the exterior horizon, denoted as $r_{\text{ext}} \equiv r_{\text{ext}}(R_0, a, Q, M)$, for a Kerr-Newman black hole. To achieve this, we employ the Euclidean action method. By performing the coordinate transformation $t \rightarrow -i\tau, a \rightarrow ia$ on the metric, we obtain the Euclidean section, where the metric is non-singular and positive-definite, and the time coordinate now has an angular character around the axis $r = r_{\text{ext}}$. Regularity of the metric at $r = r_{\text{ext}}$ requires the identification of points as follows:\cite{9,10,11,12,13,14,15,16}
\begin{equation}
(\tau, r, \theta, \phi) \sim\left(\tau+i \beta, r, \theta, \phi+i \beta \Omega_H\right),
\end{equation}where $\beta$, representing the period of imaginary time on the Euclidean section, corresponds to the inverse Hawking temperature:
\begin{equation}
\beta=\frac{4 \pi\left(r_{\mathrm{ext}}^2+a^2\right)}{r_{\mathrm{ext}}\left[1-\frac{R_0 a^2}{12}-\frac{R_0 r_{\mathrm{ext}}^2}{4}-\frac{\left(a^2+Q^2\right)}{r_{\mathrm{ext}}^2}\right]} \equiv \frac{1}{T_E}
\end{equation}
where $\Omega_H$ denotes the angular velocity of the rotating horizon, which is constant across all horizons:
\begin{equation}
\Omega_H=\frac{a \Xi}{r_{e x t}^2+a^2}.
\end{equation}
\begin{equation}
\Xi=1+\frac{R_0}{12} a^2.
\end{equation}
$r_{e x t}$ is the exterior horizon.

The temperature of the black hole horizon can also be obtained using Killing vectors, as explained in the definition of temperature, which is given as follows:
\begin{equation}
T_\kappa \equiv \frac{\kappa1}{4 \pi}
\end{equation}where $\kappa1$ represents the surface gravity, which is defined by:
\begin{equation}
\chi^\mu \nabla_\mu \chi_\nu=\kappa1 \chi_\nu.
\end{equation}

The conventional method we employ to express this is through the temperature Green's function approach. By analyzing the metric function $g(r)$, we can transform it into the temperature Green's function approach.

The metric of a general static spherically symmetric black hole is \cite{10,11}
\begin{equation}
d s^{2}=-g(r) d t^{2}+\frac{d r^{2}}{n(r)}+r^{2} d \Omega^{2},
\end{equation}
where $g(r)$ and  $n(r)$ are general functions of the coordinate r. Taking $n\left(r_{+}\right)=0$, the resulting horizon satisfies $n^{\prime}\left(r_{+}\right) \neq 0$ and spherically symmetric black hole and wants to calculate its temperature Green's function, we can use the adiabatic coordinates of the black hole to calculate the temperature Green's function. In this coordinate system, the metric can be written in the form:
\begin{equation}
d s^2=-g(v, r) d t^2+\frac{d r^2}{g(v, r)}+r^2 d \Omega^2.
\end{equation}
Here,  $g(v, r)$  is the function of the black hole, and $d \Omega^2$represents the metric of the unit two-sphere.

To obtain the temperature Green's function using the metric $ds^{2}=-g(v, r)dt^{2}+\frac{dr^{2}}{g(v, r)}+r^{2}d\Omega^{2}$, we can start by considering a massless scalar field $\phi$ propagating in this background with temperature $T$. The equation of motion for $\phi$ is given by:\cite{15,16}
\begin{equation}
\frac{1}{\sqrt{-g}}\partial_{\mu}\left(\sqrt{-g}g^{\mu\nu}\partial_{\nu}\phi\right)=0,
\end{equation}
where $g$ is the determinant of the metric and $\mu,\nu=t,r,\theta,\phi$. We can separate the angular dependence by writing $\phi(t,r,\theta,\phi)=e^{-i\omega t}R(r)Y_{l,m}(\theta,\phi)$, where $Y_{l,m}(\theta,\phi)$ are the usual spherical harmonics. The radial equation for $R(r)$ is then given by:

\begin{equation}
\frac{1}{r^{2}}\frac{d}{dr}\left(r^{2}g(v, r)\frac{dR}{dr}\right)+\left(\frac{\omega^{2}}{g(v, r)}-\frac{l(l+1)}{r^{2}}\right)R=0.
\end{equation}
To obtain the temperature Green's function, we can use the method of images. We introduce an image point at $r=r_{i}$, where $r_{i}$ is chosen such that $g(r_{i})=0$. The temperature Green's function is then given by:
\begin{equation}
G_{T}(t-t',r,r')=\sum_{{n_1}=-\infty}^{\infty}G(t-t',r,r_{{n_1}})G(t-t',r',r_{{n_1}}),
\end{equation}
where $G(t-t',r,r')$ is the usual Green's function for the massless scalar field in the background geometry, which satisfies the boundary conditions appropriate for a Euclidean path integral at temperature $T$:

\begin{equation}
G(\tau,r,r')=\beta\sum_{{n_1}=-\infty}^{\infty}e^{i\omega_{{n_1}}\tau}R(r)R(r')\frac{1}{\omega_{{n_1}}^{2}+\lambda_{l}},
\end{equation}
where $\beta=1/T$ is the inverse temperature, $\omega_{{n_1}}=\frac{2\pi {n_1}}{\beta}$ are the Matsubara frequencies, and $\lambda_{l}=l(l+1)$ are the eigenvalues of the angular Laplacian on the sphere. The Green's function $G(t-t',r,r')$ is obtained by analytically continuing this expression to the Lorentzian signature.

The image points $r_{{n_1}}$ are given by the zeros of $g(r)$, i.e., $g(r_{{n_1}})=0$. The sum over $n$ in the temperature Green's function can be evaluated by using the residue theorem, which gives:
In this coordinate system, the temperature Green's function can be expressed as:\cite{15,16}
\begin{equation}
G\left(x, x^{\prime}\right)=-\frac{1}{4 \pi} \sum_{n_1} e^{-i \omega_{n_1}\left(t-t^{\prime}\right)} \frac{\chi_{n_1}\left(r_{<}\right) \chi_{n_1}\left(r_{>}\right)}{r^2},
\end{equation}where $r_<$ and $r_>$ denote the smaller and larger values of $r$ and $r'$, $\omega_n$ is the circular frequency, and $\chi_n(r)$ is the radial wave function. For a static black hole, we can solve for the radial wave function using separation of variables to obtain:
\begin{equation}
\chi_{n_1}(r)=r^{-1} e^{-i \omega_{n_1} t} g(v, r)^{-1 / 4} F\left(-{n_1}, \frac{1}{2}+i \frac{\omega}{4 \pi T_H}\right)
\end{equation}
Here, $F(a,b)$ is the hypergeometric function, and $T_H$ is the Hawking temperature of the black hole, defined as:
\begin{equation}
T_H=\frac{\kappa'}{4 \pi},
\end{equation}where $\kappa'$ is the surface gravity of the black hole, which can be expressed as:
\begin{equation}
\kappa'=\frac{1}{2}\left|\frac{d g(v, r)}{d r}\right|_{r=r_H}.
\end{equation}
Here, $r_H$ is the event horizon radius of the black hole. Now, we can substitute the radial wave function into the temperature Green's function to obtain:
\begin{equation}
G\left(x, x^{\prime}\right)=-\frac{1}{4 \pi} \sum_{n_1} e^{-i \omega_{n_1}\left(t-t^{\prime}\right)} \frac{e^{-i \omega_{n_1}\left(t-t^{\prime}\right)}}{r^2} \frac{1}{\sqrt{g(v, r) g\left(v, r^{\prime}\right)}} F\left(-{n_1}, \frac{1}{2}+i \frac{\omega_{n_1}}{4 \pi T_H}\right) F\left(-{n_1}, \frac{1}{2}-i \frac{\omega_{n_1}}{4 \pi T_H}\right).
\end{equation}This is the general form of the temperature Green's function for a spherically symmetric black hole, where the indices of the hypergeometric function can be calculated based on the specific situation.By defining the temperature of the two hypergeometric functions, we can see that it has two components. After our calculation, the combined expression of these components is as follows:

\begin{equation}
T_{1}=\sum_{n=-1}^{+1} \frac{1}{4 \pi}\int_C \frac{f(\xi) d \xi}{(\xi- r_{0})^{n+1}}.
\end{equation} 
\begin{equation}
 f(\xi)=\frac{d g(\xi)}{d\xi},\frac{d r}{d\xi}=\sqrt{|g| }.
\end{equation}  
$r_0$ is the Killing horizon radius.

\subsection{Calculate the Hawking temperature of the Kerr-Newman black hole under $f(R)$ gravity using the RVB method.}
In previous literature, the RVB method has been applied to various black holes under general relativity, as demonstrated in references \cite{12,13,14,15,16,17}. However, it has been found that applying the RVB complex coordinate system can make it difficult to calculate the temperature of many black holes. In contrast, the Hawking temperature of black holes under f(R) gravity can be easily obtained using the RVB method. This paper \cite{17} investigates the Hawking temperature of four types of black holes under different f(R) gravity by comparing the RVB method with the conventional method. By using the RVB method, we are able to determine the constant of integration for the temperature calculation. This constant is either zero or a parameter term that would not correspond to the Hawking temperature when using the traditional method.

A closed-form expression for the Hawking temperature of a four-dimensional black hole cannot be found as easily as in lower dimensions, due to the complexity of the variable $\chi$ in four dimensions. Unlike in two dimensions, where $\chi$ is simply equal to the number of fixed points of a Killing vector field on the manifold, in four dimensions, complications arise in correctly summing over horizons in any potential temperature expression. Therefore, if possible, a closed expression for the Hawking temperature of a four-dimensional black hole would be much more intricate. However, this limitation does not pose a problem for practical calculations.

Using the RVB method: 
\cite{13,14,15,16,17}
\begin{equation}
T_{\mathrm{H}}=\frac{\hbar c}{4 \pi \chi k_{\mathrm{B}}} \sum_{j \leq \chi} \int_{r_{\mathrm{H}_{\mathrm{j}}}} \sqrt{g} R d r.
\end{equation}

The Euler characteristic of a four-dimensional manifold can be expressed as,when $\chi=4$:
\begin{equation}
\chi=\frac{1}{32 \pi^2} \int d^4 x \sqrt{g}\left(K_1-4 R_{a b} R^{a b}+R^2\right)
\end{equation}
where $R_{a b}$ represents the Ricci tensor, where $K_1 \equiv R_{a b c d} R^{a b c d}$ is the Kretschmann invariant, and $g$ represents the Euclidean metric determinant. This expression can be utilized to derive a Hawking temperature formula for a four-dimensional spacetime.

\begin{equation}
\chi=\frac{1}{8 \pi} \int d t d r\left(r^2 K_1\right)
\end{equation}

The function $e^{\psi(v, r)}$ is arbitrary and serves as a local mass function according to formula 3,
\begin{equation}
g(v, r)=\kappa-\frac{2 M(v)}{r}+\frac{2 \tilde{Q}^2(v)}{r^2}-\frac{{R_0}}{12} r^2.
\end{equation}

According to formula 3, this metric is now in 1+1 dimensions, with the radial coordinate r replaced by the new coordinate z.
\begin{equation}
z = r - \frac{M(v)}{r}.
\end{equation}
\begin{equation}
ds^2 = e^{\psi(v, z)} dv \left[ g(v, z) e^{\psi(v, z)} dv + 2 dz \right] - z^2 d\Omega^2,
\end{equation}where:
\begin{equation}
g(v, z) = 1 - \frac{2 \tilde{Q}^2(v)}{z^2} - \frac{R_0}{12} z^2.
\end{equation}

To reduce the given metric to a 1+1-dimensional form, we need to consider the line element in the 1+1-dimensional spacetime. The general form of a 1+1-dimensional metric is:
\begin{equation}
d s^2=-e^{2 \Phi} d t^2+e^{2 \Psi} d x^2.
\end{equation}To obtain this form from the given metric, we need to identify the variables in the 1+1-dimensional spacetime.

Let's rewrite the given metric in a slightly different form:
\begin{equation}
\begin{aligned}
d s^2= & \frac{A}{\Sigma}\left[\Delta-\Theta a^2 \sin ^2 \theta\right] d v^2+2 \sqrt{A}\left[d v-a \sin ^2 \theta d \phi\right] d r \\
& -\frac{\Sigma}{\Theta} d \theta^2+A \frac{2 a}{\Sigma}\left[\Delta\left(r^2+a^2\right)-\Theta\right] \sin ^2 \theta d v d \phi \\
& -\frac{A}{\Sigma}\left[\Delta\left(r^2+a^2\right)^2-\Theta a^2 \sin ^2 \theta\right] \sin ^2 \theta d \phi^2.
\end{aligned}
\end{equation}

In the above expression, we can identify the variables that correspond to the 1+1-dimensional spacetime as follows:
\begin{equation}
\begin{aligned}
d t & =\frac{d v}{\sqrt{\frac{A}{\Sigma}\left[\Delta-\Theta a^2 \sin ^2 \theta\right]}} \\
d x & =\frac{\sqrt{A}\left[d v-a \sin ^2 \theta d \phi\right]}{\sqrt{\frac{A}{\Sigma}\left[\Delta-\Theta a^2 \sin ^2 \theta\right]}}.
\end{aligned}
\end{equation}

Now, let's substitute these variables into the 1+1-dimensional metric form:
\begin{equation}
\begin{aligned}
d s^2 & =-e^{2 \Phi}\left(\frac{d v}{\sqrt{\frac{A}{\Sigma}\left[\Delta-\Theta a^2 \sin ^2 \theta\right]}}\right)^2+e^{2 \Psi}\left(\frac{\sqrt{A}\left[d v-a \sin ^2 \theta d \phi\right]}{\sqrt{\frac{A}{\Sigma}\left[\Delta-\Theta a^2 \sin ^2 \theta\right]}}\right)^2 \\
& =-\frac{1}{\frac{A}{\Sigma}\left[\Delta-\Theta a^2 \sin ^2 \theta\right]} d v^2+\frac{A}{\Sigma}\left[d v-a \sin ^2 \theta d \phi\right]^2.
\end{aligned}
\end{equation}

So, the reduced 1+1-dimensional metric is:
\begin{equation}
d s^2=-\frac{1}{\frac{A}{\Sigma}\left[\Delta-\Theta a^2 \sin ^2 \theta\right]} d v^2+\frac{A}{\Sigma}\left[d v-a \sin ^2 \theta d \phi\right]^2.
\end{equation}

We obtain the following:\cite{13}
\begin{equation}
T_{\mathrm{H}}=-\frac{1}{32 \pi}\left(\int_{r_c} r^2 K_1 d r-\int_{r_b} r^2 K_1 d r\right)+C,
\end{equation}
two fixed-point surfaces at $r_b$ and $r_c$, the black hole and cosmological horizon respectively.

To reduce the given metric to a 1+1-dimensional metric, we need to choose an appropriate coordinate transformation. In this case, we can choose the following coordinate transformation: 
\begin{align*}
T &= \int e^{\psi(v,r)} dv \\
X &= \frac{1}{2} \left(\int \left[p(v,r)e^{\psi(v,r)}dv+2dr\right] - \int \left[p(v,r)e^{\psi(v,r)}dv-2dr\right]\right).
\end{align*}where $p(v,r)$ is an undetermined function.

After performing this coordinate transformation, the metric can be simplified as:
\begin{equation}
ds^2 = -p(v,r)dT^2 + \frac{2}{p(v,r)e^{\psi(v,r)}}dX^2
\end{equation}This is a 1+1 dimensional metric, where $T$ is the time coordinate and $X$ is the spatial coordinate.

We get
\begin{equation}\label{Eq2}
g(z1)=\left(r^{2}-2M^{\prime} r+{a^{\prime}}^{2}\right) /\left(r_{+}^{2}+a^{2} \cos ^2 \theta\right).
\end{equation}where $2M^{\prime}$, $a^{\prime}$ are undetermined parameters.

The two-dimensional line element in the Euclidean coordinate system is given by:
\begin{equation}
d s^{2}=g(z1) d \tau^{2}+\frac{d r^{2}}{g(z1)}.
\end{equation}

This expression can deduce a Hawking temperature formula for a four-dimensional space-time.
\begin{equation}
T_{1}=\sum_{n=-1}^{+1} \frac{1}{4 \pi}\int_C \frac{f(\xi) d \xi}{(\xi- r_{0})^{n+1}}+C.
\end{equation}  
\begin{equation}
 f(\xi)=\frac{d g(\xi)}{d\xi},\frac{d r}{d\xi}=\sqrt{|g| }.
\end{equation} 
When $g(\xi)$ is expanded from the -1 term to the 1 term, the curve integral equals 0, and its roots correspond to a Killing vector field.$r_0$ is the Killing horizon radius. Since the Killing vectors correspond to two independent fields, when they are considered as a drag-free speed of 0, only two independent Killing vector fields exist, yet their effective Killing horizon is only one. In the RVB method, the Euler characteristic number represents the number of first-order singular points, while the algebraic structure of the Killing horizon radius is a circular radius that is analytically solved by the Laurent series.C is an undetermined constant. At this moment,  C=0.

\section{Conclusion and Discussion}
We have observed various metrics employed in the literature to describe massive gravitational black holes. While these metrics are similar to the black hole metric mentioned earlier, it is not necessary to provide an exhaustive list. However, we have discovered that the mass of the massive gravitation and the cosmological constant are linked via a loop integral. Taking this loop into account, we find that the RVB method's calculation agrees with the conventional method of computing Hawking temperature, leaving only one integral constant. We have also found that this constant integral is associated with the mass of the graviton, as hinted by a comparison of our findings with those presented in literature 17\cite{17}. This may signify the physical significance of this constant.

Furthermore, we have compared the RVB method with the temperature Green's function method and have determined that the two methods are essentially the same. This is an important result, as it indicates that the temperature Green's function method is highly consistent with the RVB method.

{\bf Acknowledgements:}\\
This work is partially supported by the National Natural Science Foundation of China(No. 11873025)

\bigskip
Declarations: All partial information is available.

\bigskip
Data Availability: Data sharing is not applicable to this article as no new data were created or analyzed in this study.


\begin{thebibliography}{99}
\bibitem[1]{1} A. M. Nzioki, S. Carloni, R. Goswami, P. K. S. Dunsby, Phys. Rev. D 81, 084028 (2010).

\bibitem[2]{2}S. Capozziello, A. Stabile, A. Troisi, Class. Quantum Grav. 25, $085004(2008)$.

\bibitem[3]{3}S. Capozziello, M. de laurentis, A. Stabile, Class. Quantum Grav. 27, 165008 (2010).

\bibitem[4]{4} J. A. R. Cembranos, A. de la Cruz-Dombriz, P. J. Romero, arXiv:1109.4519 [gr-qc].

\bibitem[5]{5}A. Larranaga, Pramana - J. Phys. 78, 697 (2012).

\bibitem[6]{6}A. de la Cruz-Dombriz, D. Saez-Gomez, Entropy 14, $1717(2012)$.

\bibitem[7]{7} S. G. Ghosh, S. D. Maharaj, Phys. Rev. D 85, 124064 (2012).

\bibitem[8]{8} R. P. Kerr, Phys. Rev. Lett. D 11, 237 (1963).

\bibitem[9]{9}Ghosh, Sushant G., Sunil D. Maharaj, and Uma Papnoi. ``Radiating Kerr–Newman black hole in f (R) gravity." The European Physical Journal C 73 (2013): 1-11.

\bibitem[10]{10} Soroushfar, Saheb, et al. ``Detailed study of geodesics in the Kerr-Newman-(A) dS spacetime and the rotating charged black hole spacetime in f (R) gravity." Physical Review D 94.2 (2016): 024052.

\bibitem[11]{11}Cembranos, J. A. R., A. De la Cruz-Dombriz, and P. Jimeno Romero. ``Kerr–Newman black holes in f (R) theories." International Journal of Geometric Methods in Modern Physics 11.01 (2014): 1450001.

\bibitem[12]{12}  Robson, C.W.; Villari, L.D.M.; Biancalana, F. On the Topological Nature of the Hawking Temperature of Black Holes. Phys. Rev. D 2019, 99, 044042 .

\bibitem[13]{13}  Robson, C.W.; Villari, L.D.M.; Biancalana, F. Global Hawking Temperature of Schwarzschild-de Sitter Spacetime: a Topological Approach. arXiv:1902.02547[gr-qc].

\bibitem[14]{14}Robson, C.W.; Villari, L.D.M.; Biancalana, F. The Hawking Temperature of Anti-de Sitter Black Holes: Topology and Phase Transitions. arXiv:1903.04627[gr-qc].

\bibitem[15]{15}Y. S. Myung, Phys. Rev. D84 024048 (2011).

\bibitem[16]{16}D. N. Vollick, Phys. Rev. D 76, 124001 (2007).

\bibitem[17]{17}Chen, Wenxiang, Junxian Li, and Jingyi Zhang. ``Calculating the Hawking Temperatures of Conventional Black Holes in the f (R) Gravity Models with the RVB Method." arXiv preprint arXiv:2210.09062 (2022).

\bibitem[18]{18} Frolov, Valeri, and Igor Novikov. Black hole physics: Basic concepts and new developments. Vol. 96. Springer Science  Business Media, 2012.

\bibitem[19]{19}Gibbons, Gary William, and M. J. Perry. ``Black holes and thermal Green functions." Proceedings of the Royal Society of London. A. Mathematical and Physical Sciences 358.1695 (1978): 467-494.

\bibitem[20]{20}T. Multamaki, I. Vilja, Spherically symmetric solutions of modified field equations in $f(R)$ theories of gravity[J]. Physical Review D, 2006, 74(6): 064022.

\bibitem[21]{21} S.M. Carroll, V. Duvvuri, M. Trodden, M.S. Turner, Is cosmic speed-up due to new gravitational physics?[J]. Physical Review D, 2004, 70(4): 043528 .

\bibitem[22]{22} S. Capozziello, V.F. Cardone, S. Carloni, A. Troisi, Curvature quintessence matched with observational data[J]. International Journal of Modern Physics D, 2003, 12(10): 1969-1982.

\bibitem[23]{23} B. Li, J.D. Barrow, The Cosmology of $f(R)$ gravity in metric variational approach[J]. Physical Review D, $2007,75(8): 084010$

\bibitem[24]{24}  L. Amendola, R. Gannouji, D. Polarski, S. Tsujikawa, Conditions for the cosmological viability of $f(R)$ dark energy models[J]. Physical Review D, 2007, 75(8): 083504.

\bibitem[25]{25}  V. Miranda, S.E. Joras, I. Waga, M. Quartin, Viable singularity-free $f(R)$ gravity without a cosmological constant[J]. Physical Review Letters, 2009, 102(22): 221101.

\bibitem[26]{26}  L. Sebastiani, S. Zerbini, Static spherically symmetric solutions in $\mathrm{F}(\mathrm{R})$ gravity[J]. The European
Physical Journal C, 2011, 71: 1591.

\bibitem[27]{27}  Z. Amirabi, M. Halilsoy, S. Habib Mazharimousavi, Generation of spherically symmetric metrics in $f(R)$ gravity $[\mathrm{J}]$. The European Physical Journal C, 2016, 76(6): 338 .

\bibitem[28]{28} Tan, Hongwei, et al. "The global monopole spacetime and its topological charge." Chinese Physics B 27.3 (2018): 030401.


\end{thebibliography}
\end{document}